\documentclass[a4paper, superscriptaddress, aps, prl, onecolumn, showkeys]{revtex4}
\usepackage{units}
\usepackage{graphicx}
\usepackage{amsmath}
\usepackage{amssymb}

\begin{document}

\author{Daniel Platz}
\email{platz@kth.se}
\author{Daniel Forchheimer}
\affiliation{Royal Institute of Technology (KTH), Section for Nanostructure Physics, Albanova University Center, SE-106 91 Stockholm, Sweden}
\author{Erik A. Thol\'{e}n}
\affiliation{Intermodulation Products AB, SE-169 58 Solna, Sweden}
\author{David B. Haviland}
\affiliation{Royal Institute of Technology (KTH), Section for Nanostructure Physics, Albanova University Center, SE-106 91 Stockholm, Sweden}

\title{Interpreting motion and force for narrow-band intermodulation atomic
force microscopy}

\begin{abstract}
Intermodulation atomic force microscopy (ImAFM) is a mode of dynamic atomic force microscopy that probes the nonlinear tip-surface force by measurement of the mixing of multiple tones in a frequency comb. A high $Q$ cantilever resonance and a suitable drive comb  will result in tip motion described by a narrow-band frequency comb. We show by a separation of time scales, that such motion is equivalent to rapid oscillations at the  cantilever resonance with a slow amplitude and phase or frequency modulation. With this time domain perspective we analyze single oscillation cycles in ImAFM to extract the Fourier components of the tip-surface force that are in-phase with tip motion ($F_I$) and quadrature to the motion ($F_Q$).  Traditionally, these force components have been considered as a function of the static probe height only. Here we show that $F_I$ and $F_Q$ actually depend on both static probe height and oscillation amplitude.  We demonstrate on simulated data how to reconstruct the amplitude dependence of $F_I$ and $F_Q$ from a single ImAFM measurement.  Furthermore, we introduce ImAFM approach measurements with which we reconstruct the full amplitude and probe height dependence of the force components $F_I$ and $F_Q$ , providing deeper insight into the tip-surface interaction. We demonstrate the capabilities of ImAFM approach measurements on a polystyrene polymer surface.
\end{abstract}


\maketitle

\section{Introduction}

\begin{figure}
\centering{}\includegraphics{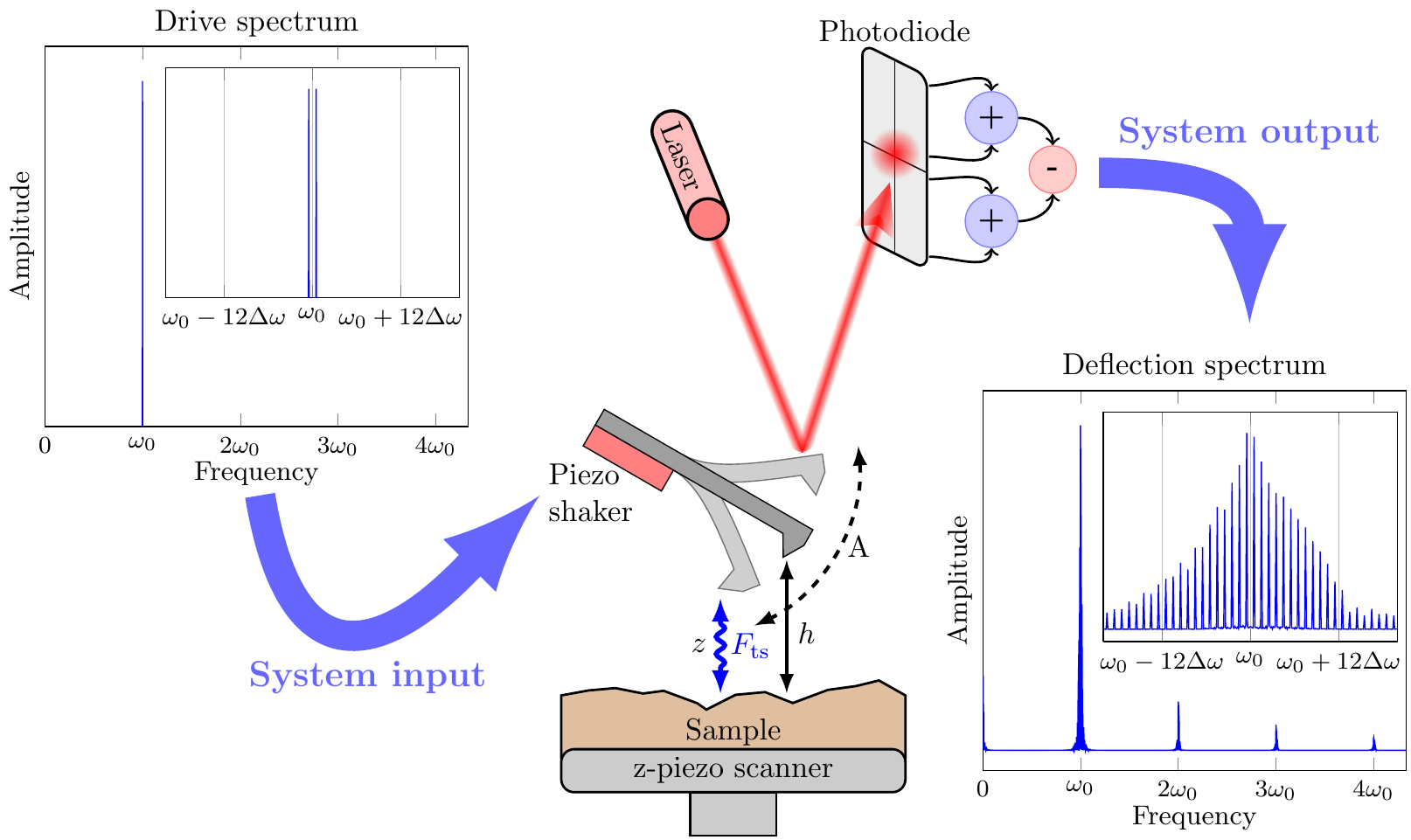}\caption{Sketch of the basic experimental setup in narrow-band ImAFM. In the absence of a drive signal the tip is at rest at the static probe height $h$.  The spectrum of the drive comb consist of two frequency components spaced by $\Delta\omega$ and centered at the first flexural resonance frequency $\omega_0$ of the cantilever which is much higher than the comb base frequency $\Delta\omega$ (here $\omega_0 = 600 \Delta \omega$). The driven tip oscillates with amplitude $A$ and interacts with the sample surface. The instantaneous tip position $z$ is measured in the rest frame of the sample surface. The corresponding deflection signal is detected by an optical lever system and is concentrated to a narrow band around $\omega_0$, as the drive signal. In this band new frequency components spaced by $\Delta\omega$ are present which are generated by the nonlinear tip-sample interaction $F_{\mathrm{ts}}$. Outside the narrow band at $\omega_0$ there is only little response in bands at integer multiples of $\omega_0$.
\label{fig:exp-setup}}
\end{figure}

Since its invention\cite{Binnig1986} atomic force microscopy (AFM) has developed into one of the most versatile techniques in surface science. At length scales ranging from micrometers down to the level of single atoms, AFM-based techniques are used to image\cite{Ohnesorge1993,Giessibl1995,Hansma1994}, measure\cite{Rugar1992,Butt2005b} and manipulate mater\cite{Perez-Murano1995,Garca1999,Sugimoto2005} at an interface. As an imaging tool, the goal of AFM development has been increasing spatial resolution and minimizing the back action force from the probe on the sample surface. A major advancement in this regard was the development of dynamic AFM\cite{Martin1987} in which a sharp tip at the free end of the AFM cantilever oscillates close to the sample surface as depicted in figure \ref{fig:exp-setup}. In order to achieve stable oscillatory motion, an external drive force is applied to the cantilever which is usually purely sinusoidal in time with a frequency that is close to the resonance frequency of the first flexural eigenmode of the cantilever. The high quality factor of the resonance ensures that the responding motion of the tip is approximately sinusoidal in time, with the same frequency as the drive signal\cite{Cleveland1998, Salapaka2000}. Such periodic motion is best analyzed in the frequency or Fourier domain, where the motion is well described by one complex-valued Fourier coefficient at the drive frequency. This motion has a corresponding Fourier coefficient of the tip-surface force, which can be expressed in terms of two real-valued components, $F_I$ which is in-phase with the motion and  $F_Q$ which is quadrature to the motion. At a fixed probe height $h$ above the surface, the two force quadratures $F_I$ and $F_Q$ give only qualitative insight into the interaction between the tip and the surface\cite{Paulo2002} and most quantitative force reconstruction methods are based on a measurement of $F_I$ and $F_Q$ at different $h$ \cite{Durig2000,Sader2005,Holscher2006,Lee2006,Hu2008,Katan2009}.

In order to increase the accessible information while imaging with AFM, a variety methods have been put forward where amplitude and phase at more than one frequency is analyzed. These multifrequency methods can be divided in to two general groups: Those using only Fourier components with frequencies close to a cantilever resonance, and those which use off-resonance components. Off-resonance techniques typically measure higher harmonics of the tip motion which allows for a reconstruction of time-dependent surface forces acting on the tip. Due to the lack of transfer gain off resonance, these off-resonance components have small signal-to-noise ratio and their measurement requires special cantilevers\cite{Sahin2007}, high interaction forces\cite{Stark2002} or highly damped environments\cite{Legleiter2006}. To increase the number of Fourier components with good signal-to-noise ratio, on-resonance techniques utilize multiple eigenmodes of the cantilever\cite{Rodrguez2004,Solares2010,Kawai2009,Platz2008}. However, accurate calibration of higher cantilever modes remains complicated since additional knowledge about the cantilever is required. Both on and off-resonance techniques require broad-band detection of the cantilever motion, which implies a sacrifice in the sensitivity and gain of the motion detection system. 

To mitigate these problems, we have developed narrow-band intermodulation AFM (ImAFM) which analyzes the response only near the first flexural eigenmode. In general ImAFM utilizes frequency mixing due to the nonlinear tip-surface interaction. A drive signal that comprises multiple  frequency components is used for exciting the cantilever, which will exhibit response not only at the drive frequencies, but also at frequencies that are linear integer combinations of the drive frequencies, 
\begin{equation}
\omega_{\mathrm{imp}}=m_{1}\omega_{1}+m_{2}\omega_{2}+\ldots+m_{M}\omega_{M}
\end{equation}
where $m_{1},m_{2},\ldots m_{M}\in\mathbb{Z}$ and $\omega_{1},\omega_{2},\ldots,\omega_{M}$ are the drive frequencies. These new frequency components are called intermodulation products (IMPs) and one usually defines an order for each IMP which is given by $|m_{1}|+|m_{2}|+\ldots+|m_{M}|$. If all frequencies in a signal are integer multiples  of a base frequency $\Delta \omega$ the signal is called a frequency comb. The nonlinear tip-surface interaction  maps a drive frequency comb to a response frequency comb, both having the same base  frequency $\Delta\omega$. Different drive frequency combs can be used to place many response frequency components close to a resonance of the cantilever where they can be detected with good signal-to-noise ratio. In general  the drive and response frequency combs could encompass more than one  eigenmode of the cantilever. For a drive signal consisting of only two frequencies symmetrically placed around the first flexural resonance frequency as illustrated in figure \ref{fig:exp-setup} the response is concentrated in the narrow band around the first resonance for which well accurate  calibration methods exist\cite{Hutter1993,Sader1999,Higgins2006}.

In what follows, we will focus on this particular case which we call narrow-band ImAFM.  However, we want to emphasize that drive schemes which generate response in more than one frequency band are also possible. We have previously shown how the individual amplitudes\cite{Platz2008} and phases\cite{Platz2010} of the IMPs in the narrow band around the first flexural resonance can be used for imaging. Furthermore, a polynomial reconstruction of the tip-surface force\cite{Hutter2010,Platz2012} and a numerical fit of  the parameters of a force model\cite{Forchheimer2012} are possible by analysis of the data in the frequency domain. Here, we consider the meaning of the narrow-band intermodulation response comb in the time domain, which leads to a physical interpretation of the intermodulation spectrum in terms of the in-phase force component $F_I$ and the quadrature force component $F_Q$.

\section{Results and discussion}

\subsection{Time domain interpretation of narrow-band frequency comb}

\begin{figure}
\centering{}\includegraphics{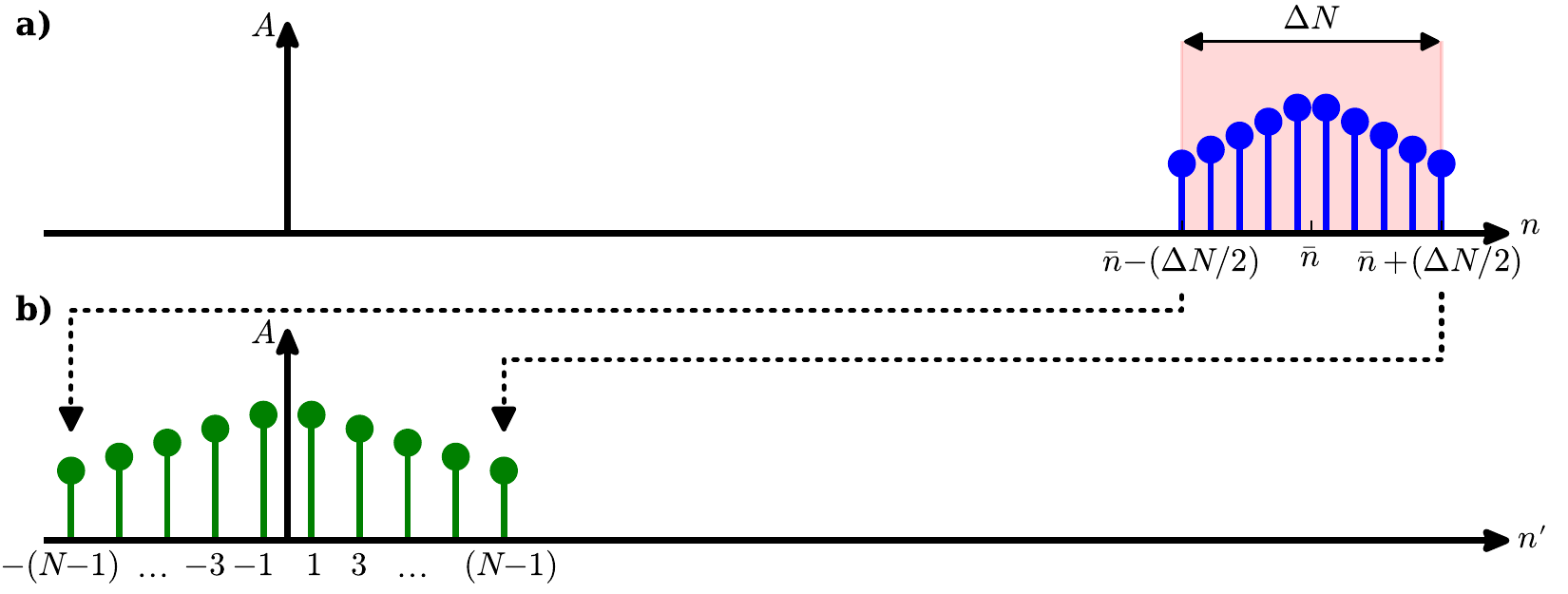}\caption{The amplitude spectrum of a narrow band signal as a function of the Fourier index(a) is characterized by a finite number of Fourier components in a frequency band around the center frequency $\bar{\omega}=\bar{n}\Delta\omega$ in a bandwidth that is given by the integer number $\Delta N$. The spectrum of the corresponding time-dependent envelope function(b) is obtained by down-shifting the original spectrum in frequency space such that the shifted center frequency is zero. \label{fig:downshift-sketch}}
\end{figure}

Figure \ref{fig:downshift-sketch}(a) portrays the amplitudes of the components of a narrow band frequency comb. While we are plotting only the amplitude at each component, it is understood that each component also has a phase. The frequency comb is characterized by the center frequency $\bar{\omega}$ of the comb, a base frequency $\Delta\omega$ and a finite number $N$ of Fourier components at discrete frequencies. Without loss of generality, we assume $N$ to be even. The center frequency $\bar{\omega}$ can be described in terms of the ratio $\bar{n}=\nicefrac{\bar{\omega}}{\Delta\omega}$ and the bandwidth of the comb is given by $\Delta N=N-1$. The $N$ discrete frequencies in the band are represented by an integer frequency index $n$ such that
\begin{equation}
\omega_{n}=n\Delta\omega
\end{equation}
where $n$ takes consecutive integer values between $\bar{n} - \nicefrac{\Delta N}{2}$ and $\bar{n} + \nicefrac{\Delta N}{2}$. In the time domain the corresponding real-valued signal $x(t)$ is then given by the Fourier series
\begin{equation}
x(t)=\sum_{n=\bar{n} - \nicefrac{\Delta N}{2}}^{\bar{n} + \nicefrac{\Delta N}{2}}\hat{x}_{n}e^{in\Delta\omega t}+\sum_{n=\bar{n} - \nicefrac{\Delta N}{2}}^{\bar{n} + \nicefrac{\Delta N}{2}}\hat{x}_{n}^{*}e^{-in\Delta\omega t}\label{eq:signal-fourier-series}
\end{equation}
where $\hat{x}_{n}$ are the complex Fourier components in the narrow frequency band and the star denotes complex conjugation. The center frequency $\bar{\omega}$ is usually much bigger than the base frequency $\Delta\omega$,
\begin{equation}
\bar{\omega}\gg\Delta\omega.
\end{equation}
Therefore, the time domain signal $x(t)$ exhibits two different time scales: a fast time scale $T_{\mathrm{fast}}=\nicefrac{2\pi}{\bar{\omega}}$ and a slow time scale $T_{\mathrm{slow}}=\nicefrac{2\pi}{\Delta\omega}$. To separate these two time scales we factor out a rapidly oscillating term at the frequency $\bar{\omega}$ from the Fourier series in equation (\ref{eq:signal-fourier-series}),
\begin{eqnarray}
x(t) & = & \sum_{n=\bar{n} - \nicefrac{\Delta N}{2}}^{\bar{n} + \nicefrac{\Delta N}{2}}\hat{x}_{n}e^{i(n\Delta\omega-\bar{\omega})t}e^{i\bar{\omega t}}+\sum_{n=\bar{n} - \nicefrac{\Delta N}{2}}^{\bar{n} + \nicefrac{\Delta N}{2}}\hat{x}_{n}^{*}e^{-i(n\Delta\omega-\bar{\omega})t}e^{-i\bar{\omega t}},\\
 & = & \sum_{n=\bar{n} - \nicefrac{\Delta N}{2}}^{\bar{n} + \nicefrac{\Delta N}{2}}\hat{x}_{n}e^{i(2n-2\bar{n})\frac{\Delta\omega}{2}t}e^{i\bar{\omega t}}+\sum_{n=\bar{n} - \nicefrac{\Delta N}{2}}^{\bar{n} + \nicefrac{\Delta N}{2}}\hat{x}_{n}^{*}e^{-i(2n-2\bar{n})\frac{\Delta\omega}{2}t}e^{-i\bar{\omega t}}.
\end{eqnarray}
Since $N$ is even, $2\bar{n}$ is an odd integer number and we can define a new sum index 
\begin{equation}
n'\equiv2n-2\bar{n}
\end{equation}
which increase in steps of $\Delta n'=2$ and the summation limits become
\begin{eqnarray}
n'_{1}  = &  2\left(\bar{n} - \frac{\Delta N}{2}\right) - 2\bar{n} & =  -(N-1),\\
n'_{N}  = & 2\left(\bar{n} + \frac{\Delta N}{2}\right) - 2\bar{n} & =  N-1.
\end{eqnarray}
Since $N$ is even and $n'$ increases only in steps of two, $n'$ can only take odd values. Additionally, we define new Fourier coefficients
\begin{equation}
\hat{x}'_{n'}\equiv\hat{x}_{\frac{1}{2}(n'+2\bar{n})}
\end{equation}
such that the signal Fourier series becomes
\begin{equation}
x(t)=\left(\sum_{\begin{array}{c}
n'=-(N-1)\\
n'\ \mathrm{odd}
\end{array}}^{N-1}\hat{x}'_{n'}e^{in'\frac{\Delta\omega}{2}t}\right)e^{i\bar{\omega}t}+\left(\sum_{\begin{array}{c}
n'=-(N-1)\\
n'\ \mathrm{odd}
\end{array}}^{N-1}\hat{x}_{n'}^{'*}e^{-in'\frac{\Delta\omega}{2}t}\right)e^{-i\bar{\omega}t}.
\end{equation}
We identify the terms in parentheses as the Fourier series of a complex-valued time-dependent envelope function $\hat{E}(t)$  expanded in the base frequency $\nicefrac{\Delta\omega}{2}$,
\begin{equation}
\hat{E}(t)=\sum_{\begin{array}{c}
n'=-(N-1)\\
n'\ \mathrm{odd}
\end{array}}^{N-1}\hat{x}'_{n'}e^{in'\frac{\Delta\omega}{2}t}\label{eq:envelope-function}
\end{equation}
 and write the original signal $x(t)$ as
\begin{equation}
x(t)=\hat{E}(t)e^{i\bar{\omega}t}+\hat{E}^{*}(t)e^{-i\bar{\omega}t}.\label{eq:signal-with-cx-envelope}
\end{equation}
The envelope function $\hat{E}(t)$ was obtained by down-shifting the narrow intermodulation frequency band to a center frequency of zero (see figure \ref{fig:downshift-sketch}). If the maximum frequency in the Fourier series of $\hat{E}(t)$ is much smaller than $\bar{\omega}$, the envelope function $\hat{E}(t)$ varies slowly compared to the term $e^{i\bar{\omega}t}$ in equation (\ref{eq:signal-with-cx-envelope}). When we represent $\hat{E}(t)$ by a time-dependent amplitude $A(t)$ and a time-dependent phase $\phi(t)$ such that 
\begin{equation}
\hat{E}(t)=A(t)e^{i\phi(t)},
\end{equation}
the signal $x(t)$ is completely described by a modulated oscillation amplitude and a modulated oscillation phase,
\begin{eqnarray}
x(t) & = & A(t)e^{i\phi(t)}e^{i\bar{\omega}t}+A(t)e^{-i\phi(t)}e^{-i\bar{\omega}t}=2A(t)\cos\left(\bar{\omega}t+\phi(t)\right).
\end{eqnarray}

We would like to emphasize that the narrow-band frequency comb can also describe amplitude and frequency modulated signals. For frequency modulation we define an instantaneous oscillation phase 
\begin{equation}
\theta(t)=\bar{\omega}t+\phi(t)
\end{equation}
and an instantaneous oscillation frequency
\begin{equation}
\omega(t)=\frac{d\theta}{dt}=\bar{\omega}+\frac{d\phi}{dt}.
\end{equation}
The instantaneous frequency shift $\delta\omega$ compared to $\bar{\omega}$ is then simply 
\begin{equation}
\delta\omega(t)=\frac{d\phi}{dt}.
\end{equation}
In a small region around the time $t'$ the signal $x(t)$ can be obtained by a Taylor expansion to first order of the instantaneous
phase $\theta$
\begin{equation}
x(t)=A(t)\cos\left(\theta(t)\right)\approx A(t)\cos\left(\left(\bar{\omega}+\delta\omega(t')\right)t+\left(\phi(t')-\delta\omega(t')t'\right)\right).
\end{equation}
Thus, a narrow-band frequency comb can describe signals with frequency shifts that are periodic in time.

\begin{figure}
\centering{}\includegraphics{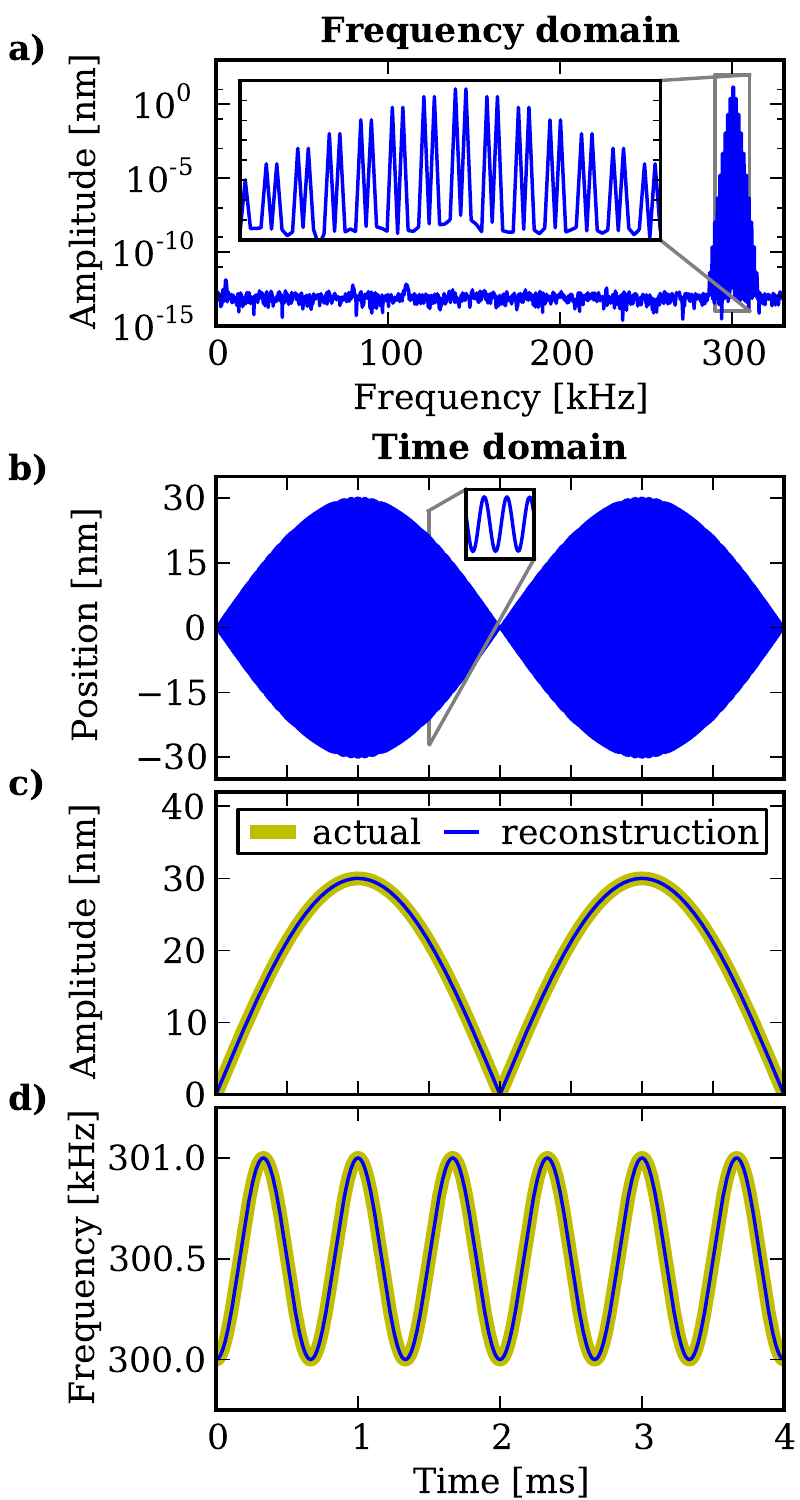}\caption{The amplitude spectrum of an amplitude and frequency modulated signal(a). The response is concentrated in a narrow-band frequency comb with a center frequency much higher than the comb base frequency. In the time domain(b) the signal rapidly oscillates on a short time scale and the slow amplitude modulation is clearly visible. The time-dependent amplitude(c) and frequency(d) reconstructed from the envelope function are in excellent agreement with the actual modulation used for the signal generation in the time domain. \label{fig:envelope-example}}
\end{figure}

To illustrate the complete description of a narrow-band signal by its envelope function, figure \ref{fig:envelope-example} shows the spectrum of an artificially constructed signal with sinusoidally modulated amplitude and frequency. Typical parameters from AFM experiments have been chosen for the amplitude and frequency modulation. The spectrum of the signal shown in figure \ref{fig:envelope-example} shows significant amplitudes at discrete frequencies in only a narrow band around 300 kHz. We down-shift the spectrum to determine the slowly varying envelope function from which we compute the time-dependent oscillation amplitude and frequency, both of which are in excellent agreement with the actual amplitudes and frequencies used for the signal generation.

To summarize, we have introduced a time domain interpretation of narrow-band frequency combs. If the center frequency of the band is much higher than the base frequency of the comb, we can separate a fast and a slow time scale in the time domain. On the fast time scale the signal rapidly oscillates at the center frequency. The slow time scale dynamics is given by the down-shifted intermodulation spectrum which describes a slow amplitude modulation and a slow phase or frequency modulation of the signal in the time domain.

\subsection{Physical interpretation of the tip motion and force envelope functions in ImAFM}

In ImAFM the measured frequency comb corresponds to a vertical motion $z(t)$ of the tip which undergoes rapid oscillations at frequency $\bar{\omega}$ with slowly varying amplitude and phase,
\begin{equation}
z(t)=A(t)\cos\left(\bar{\omega}t+\phi(t)\right)+h
\end{equation}
where $h$ is the static probe height above the surface and the amplitude $A(t)$ and the phase $\phi(t)$ are determined from the complex-valued motion envelope function $\hat{E}_{z}(t)$ as 
\begin{eqnarray}
A(t) & = & \left|\hat{E}_{z}(t)\right|,\\
\phi(t) & = & \arg\left(\hat{E}_{z}(t)\right).
\end{eqnarray}
The envelope function $\hat{E}_{z}(t)$ was obtained directly from the measured motion spectrum using equation (\ref{eq:envelope-function}).

\begin{figure}
\begin{centering}
\includegraphics{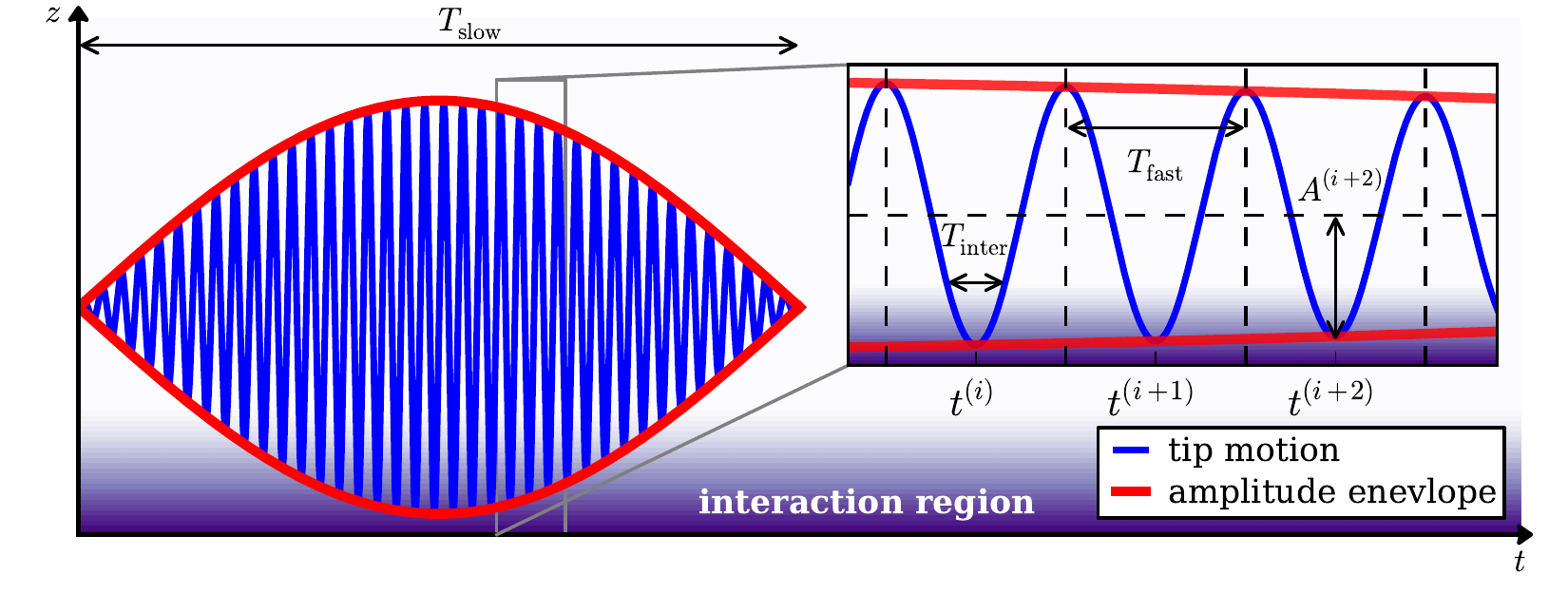}
\par\end{centering}

\caption{Sketch of a narrow band signal. On the slow time scale $T_{\mathrm{slow}}$ the tip motion shows an amplitude modulation. On the fast time scale $T_{\mathrm{fast}}$ the signal rapidly oscillates. During each oscillation cycle the tip interacts with the surface for the time $T_{\mathrm{inter}}$ during which the oscillation amplitude phase are approximately constant.\label{fig.sketch-motion-decomposition}}
\end{figure}

Knowledge of the cantilever transfer function $\hat{G}$ and the applied drive force allows for converting the measured motion spectrum into the spectrum of the time-dependent tip-surface force acting on the tip. However, the force spectrum is incomplete since higher frequency components of the force are filtered out from the motion spectrum by the sharply peaked cantilever transfer function. The time-dependence of the corresponding partial force signal is described by the force envelope function $\hat{E}_{F}(t)$,
\begin{equation}
F_{\mathrm{partial}}(t)=\hat{E}_{F}(t)e^{i\bar{\omega}t}+\hat{E}_{F}^{*}(t)e^{-i\bar{\omega}t}
\end{equation}
where $\hat{E}_{F}(t)$ is determined by applying equation (\ref{eq:envelope-function}) to the partial force spectrum. To understand the physical meaning of the partial force we analyze the signals at the level of single rapid oscillation cycles in the time domain. During each oscillation cycle the tip interacts with the sample surface. This interaction is very localized (a few nanometers above the surface) compared to the oscillation amplitude (tens of nanometers) and thus the interaction time $T_{\mathrm{inter}}$ is short compared to the fast oscillation period $T_{\mathrm{fast}}$ which itself is much shorter than the period of the beat
$T_{\mathrm{slow}}$ (see figure \ref{fig.sketch-motion-decomposition}),
\begin{equation}
T_{\mathrm{inter}}< T_{\mathrm{fast}}\ll T_{\mathrm{slow}}
\end{equation}
Therefore, amplitude $A(t)$, phase $\phi(t)$ and force envelope function $\hat{E}_{F}(t)$ can be considered to be constant during each interaction cycle and the motion and the partial force during the $i$-th tip oscillation cycle are given by
\begin{eqnarray}
z^{(i)}(t) & = & A^{(i)}\cos\left(\bar{\omega}t+\phi^{(i)}\right)+h\\
F_{\mathrm{partial}}^{(i)}(t) & = & \hat{E}_{F}^{(i)}e^{i\bar{\omega}t}+\hat{E}_{F}^{(i)*}e^{-i\bar{\omega}t}\label{eq:partial-force-fourier}
\end{eqnarray}
where $A^{(i)}$, $\phi^{(i)}$ and $\hat{E}_{F}^{(i)}$ are constant and are determined at the time $t^{(i)}$ of the ith lower turning point of the tip motion
\begin{eqnarray}
A^{(i)} & = & A\left(t^{(i)}\right)=\left|\hat{E}_{z}\left(t^{(i)}\right)\right|\\
\phi^{(i)} & = & \phi\left(t^{(i)}\right)=\arg\left(\hat{E}_{z}\left(t^{(i)}\right)\right)\\
\hat{E}_{F}^{(i)} & = & \hat{E}_{F}\left(t^{(i)}\right)
\end{eqnarray}
The complete time-dependent tip-surface force during an interaction cycle is a force pulse which can be written as a Fourier series in the oscillation frequency $\bar{\omega}$ as 
\begin{equation}
F_{\mathrm{complete}}^{(i)}(t)=\sum_{n=-\infty}^{\infty}\hat{F}_{n}^{(i)}e^{in\bar{\omega}}\label{eq:complete-force-fourier}
\end{equation}
where the complex Fourier components $\hat{F}_{n}^{(i)}$ fulfill the relation $\hat{F}_{-n}^{(i)}=\hat{F}_{n}^{(i)*}$. Comparison of equation (\ref{eq:partial-force-fourier}) with equation (\ref{eq:complete-force-fourier}) yields
\begin{equation}
\hat{F}_{1}^{(i)}=\hat{E}_{F}^{(i)}=\hat{E}_{F}\left(t^{(i)}\right)
\end{equation}
which reveals that the first Fourier component $\hat{F}_{1}^{(i)}$ of the force pulse during the ith oscillation cycle is given by the force envelope function $\hat{E}_{F}$ determined from the partial force spectrum. Since each lower motion turning point is associated with a unique amplitude $A^{(i)}$ we can consider $\hat{F}_1$ as a function of the continuous variable $A$,
\begin{equation}
\hat{F}^{(i)}_{1}=\hat{F}_{1}(A^{(i)}) \Rightarrow \hat{F}_1 = \hat{F}_1(A).
\end{equation}
The amplitude-dependence of $\hat{F}_{1}$ can then be uncovered by the analysis of all oscillation cycles during the time $T_{\mathrm{slow}}$. 

To better relate motion and the force we compute the components of $\hat{F}_{1}(A)$ that are in phase with the motion ($F_I$) and quadrature to the motion ($F_Q$). For a tip-surface force that only depends on the instantaneous tip position and velocity,
\begin{equation}
F_{\mathrm{ts}}=F_{\mathrm{ts}}(z,\dot{z}),
\end{equation}
we approximate the tip motion to be purely sinusoidal at frequency
$\bar{\omega}$ with amplitude $A$ and without an additional phase.
At fixed probe height $h$, the components $F_I$ and $F_Q$
are given by two integral equations
\begin{eqnarray}
F_I(A,h) & = & \frac{1}{T_{\mathrm{fast}}}\int_{0}^{T_{\mathrm{fast}}}F_{\mathrm{ts}}\left(A\cos(\bar{\omega}t)+h,-\bar{\omega}A\sin(\bar{\omega}t)\right)\cos(\bar{\omega}t)dt\label{eq:FI-int}\\
F_Q(A,h) & = & \frac{1}{T_{\mathrm{fast}}}\int_{0}^{T_{\mathrm{fast}}}F_{\mathrm{ts}}\left(A\cos(\bar{\omega}t)+h,-\bar{\omega}A\sin(\bar{\omega}t)\right)\sin(\bar{\omega}t)dt\label{eq:FQ-int}
\end{eqnarray}
With these assumptions $F_I$ becomes the so-called virial of the tip-surface force which is only affected by the conservative part of the tip-surface interaction\cite{Paulo2001} whereas $F_Q$ is related to the energy dissipated by the tip-surface interaction\cite{Cleveland1998}. We note that through their dependence on tip position $z$ and velocity $\dot{z}$, the force components $F_I$ and $F_Q$ are functions of both probe height $h$ and oscillation amplitude $A$ . However, usually they are considered as functions of the probe height $h$ only.

For an ImAFM measurement at fixed probe height, the amplitude-dependence of $F_I$ and $F_Q$ can readily be obtained by defining a new force envelope function that is phase-shifted with respect to the motion by the angle $\phi$, 
\begin{equation}
\hat{E}'_{F}(t)=\hat{E}_{F}(t)e^{-i\phi(t)}
\end{equation}
which we evaluate as real and imaginary parts at the times of the lower turning points of the motion 
\begin{eqnarray}
F_I(A^{(i)}) & = & \mathrm{Re}\left(\hat{E}'_{F}(t^{(i)})\right),\label{eq:FI-EF}\\
F_Q(A^{(i)}) & = & \mathrm{Im}\left(\hat{E}'_{F}(t^{(i)})\right)\label{eq:FQ-EF}
\end{eqnarray}
With this interpretation of an intermodulation spectrum we are able to reconstruct the amplitude-dependence of the force quadratures $F_I$ and $F_Q$ which are independent of details of the tip motion on the slow time scale. Due to this independence, $F_I$ and $F_Q$ are the input quantities for nearly all force spectroscopy methods in dynamic AFM and thereby they form the basis of quantitative dynamic AFM.

\subsection{Force quadrature reconstruction from simulated data}

\begin{figure}
\centering{}\includegraphics{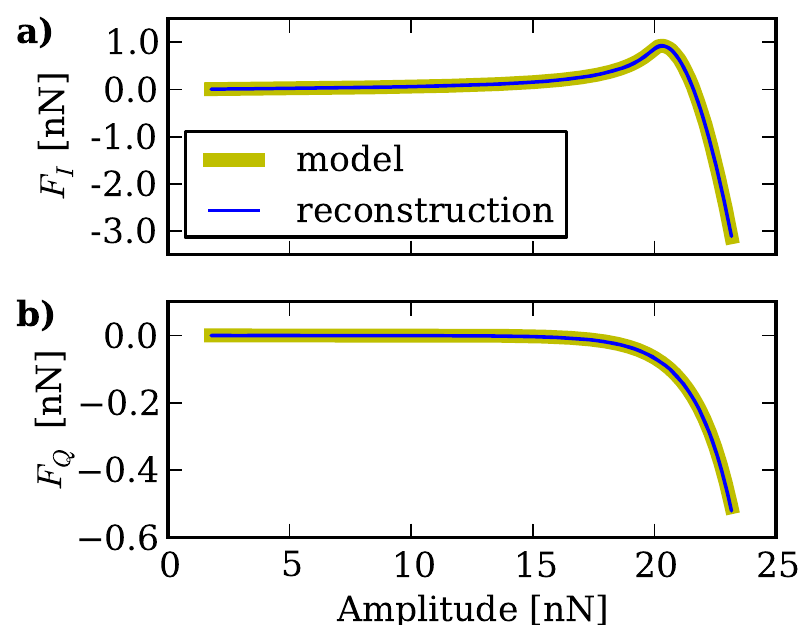}\caption{The $F_I(A)$ and $F_Q(A)$ curves reconstructed from simulated tip motion in ImAFM. The reconstructed curves  are in good agreement with the actual curves directly determined from the model force used in the simulations.\label{fig:F-sim-recon}}
\end{figure}

To demonstrate the accuracy of the $F_I$(A) and $F_Q(A)$ reconstruction from ImAFM data we simulate the tip motion in a model force field. We excite the tip with two frequencies close to the first flexural resonance frequency of $\omega_{0}=2\pi\cdot300\ \mathrm{kHz}$ which allows us to model the cantilever as a single eigenmode system for which the tip dynamics are described by an effective harmonic oscillator equation\cite{Rodrguez2002,Melcher2007}
\begin{equation}
\ddot{z}+\frac{\omega_{0}}{Q}\dot{z}+\omega_{0}^{2}(z-h)=\frac{\omega_{0}^{2}}{k_c}\left(F_{1}\cos(\omega_{1}t)+F_{2}\cos(\omega_{2}t)+F_{\mathrm{ts}}(z,\dot{z})\right)\label{eq:equation-of-motion}
\end{equation}
where $Q=400$ is the quality factor of the resonance, $k_c=40\ \mathrm{N/m}$ is the mode stiffness and $h=20\ \mathrm{nm}$ is the static probe height above the surface. The drive strengths $F_{1}$ and $F_{2}$ at the frequencies $\omega_{1}=2\pi\cdot299.75\ k\mathrm{Hz}$ and $\omega_{2}=2\pi\cdot300.25\ k\mathrm{Hz}$ are chosen such that in the absence of a tip-surface force, the tip oscillation amplitude is sinusoidally modulated between 0 nm and 30 nm . For the tip-surface force $F_{\mathrm{ts}}$ we assume a van-der-Waals-Derjaguin-Muller-Toporov (vdW-DMT) force with additional exponential damping as  which is defined as
\begin{equation}
\label{eq:tip-surface-force} 
F_{\mathrm{ts}}(z,\dot{z})=
\begin{cases}
-\frac{HR}{6(a_{0}+z)^{2}}-\gamma\exp\left(-\nicefrac{z}{z_{\gamma}}\right)\dot{z} & z\ge0\\
-\frac{HR}{6a_{0}^{2}}+\frac{4}{3}\sqrt{R}E^{*}(-z)^{3/2}-\gamma\exp\left(-\nicefrac{z}{z_{\gamma}}\right)\dot{z} & z<0
\end{cases}
\end{equation}
where $H=2.96\cdot10^{-7}\mathrm{\ J}$ is the Hamaker constant, $R=10\ \mathrm{nm}$ is the tip radius, $\gamma=2.2\cdot10^{-7}\ \mathrm{Ns/m}$ is the damping constant, $z_{\gamma}=1.5\ \mathrm{nm}$ is the damping decay length and $E^{*}=2.0\ \mathrm{GPa}$ is the effective stiffness. For the numerical integration of equation (\ref{eq:equation-of-motion}) we use the adaptive step-size integrator cvode\cite{Hindmarsh2005} with root detection to properly treat the piecewise definition of the tip-surface force in equation (\ref{eq:tip-surface-force}). From the simulated tip motion we determine the motion and the force envelope functions $\hat{E}_{z}$ and $\hat{E}_{F}$ and reconstruct $F_I$(A) and $F_Q(A)$ according to equations (\ref{eq:FI-EF}) and (\ref{eq:FQ-EF}). As shown in figure (\ref{fig:F-sim-recon}) the reconstructed curves are in excellent agreement with the curves directly computed with equations (\ref{eq:FI-int}) and (\ref{eq:FQ-int}) from the model force used in the simulations.

\subsection{Probing the force quadratures}

\begin{figure}
\begin{centering}
\includegraphics{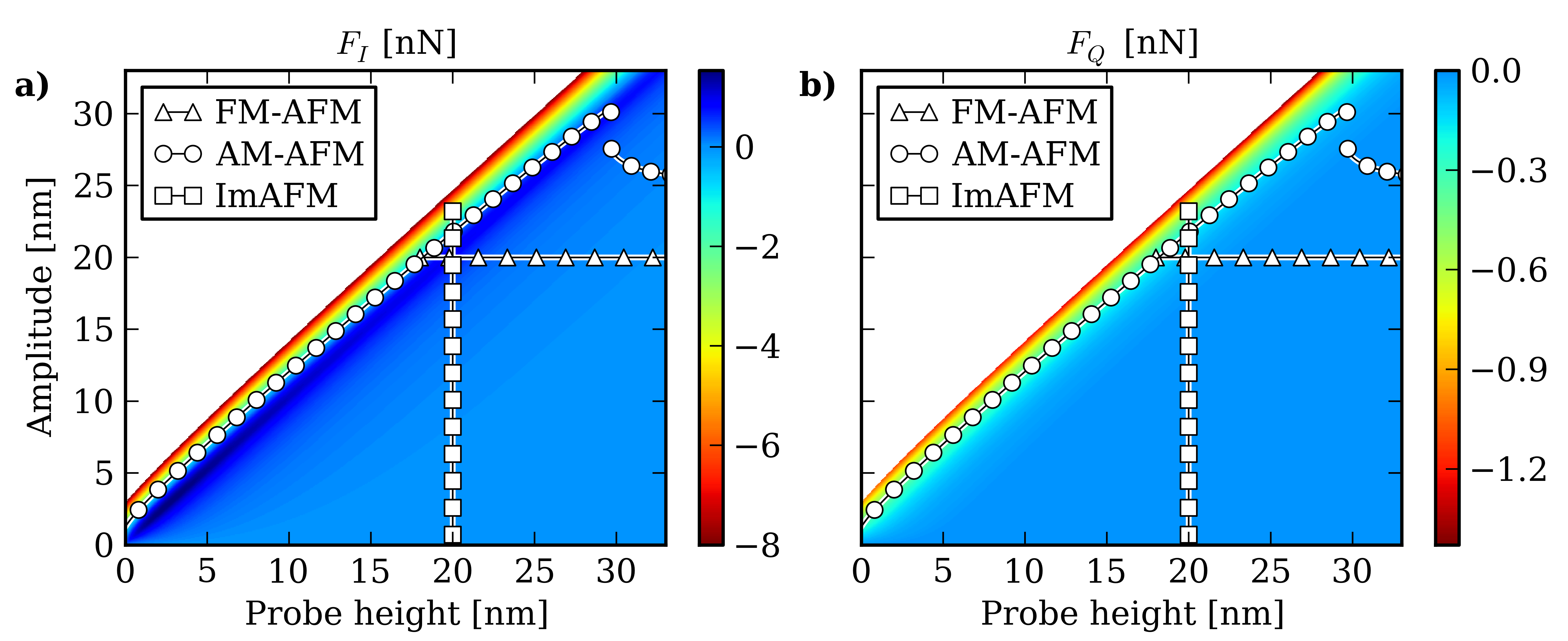}
\par\end{centering}

\caption{Model $F_I(h,A)$ and $F_Q(h,A)$ maps for the vdW-DMT force with exponential damping introduced in equation (\ref{eq:tip-surface-force}). The displayed measurement paths correspond to a frequency-shift-distance curve in FM-AFM, an amplitude-phase-distance curve in AM-AFM and an ImAFM measurement. In contrast to FM-AFM and AM-AFM, the static probe height is constant during an ImAFM measurement and the $h$-$A$ plane is explored along a path parallel to the $A$ axis. One should also note the amplitude jump along the AM-AFM path at a probe height of $h=30\ \mathrm{nm}$.\label{fig:F-model-maps}}
\end{figure}

The force quadratures $F_I$ and $F_Q$ are the basic input quantities for a variety of force reconstruction techniques\cite{Durig2000,Sader2005,Holscher2006,Lee2006,Hu2008,Katan2009}. Over the last decade the dominate paradigm was to consider $F_I$ and $F_Q$ as functions of the static probe height \textsl{$h$} only, and only one oscillation amplitude $A$ was considered at each probe height. $F_I$ and $F_Q$ are however functions of both $h$ and $A$ as seen in the two-dimensional color maps shown in figure \ref{fig:F-model-maps} for the vdw-DMT force with exponential damping used in the previous section. In order to emphasize the interaction region near the point of contact, data in the $h$-$A$ plane with $F_{I}<-8\ \mathrm{nN}$ are masked with white.

In both frequency modulation AFM (FM-AFM) and amplitude modulation AFM (AM-AFM) $F_I$ and $F_Q$ are usually probed by a slow variation of the probe height $h$ with fixed oscillation amplitude at each height. To measure $F_I$ and $F_Q$ in FM-AFM the oscillation frequency shift and the drive force are recorded as the static probe height is slowly varied (frequency-shift-distance curves). Active feedback is used to adjust both the drive power and drive frequency, as to keep the response amplitude and phase constant. The obtained frequency shifts and drive forces can then be converted into the force quadratures\cite{Giessibl1997,Giessibl2002} so that the measurement corresponds to a measurement of $F_I$ and $F_Q$ along a path parallel to the $h$-axis in the $h$-$A$-plane (see figure \ref{fig:F-model-maps}).

In AM-AFM the oscillation amplitude and phase with respect to the drive force are measured as a function of the static probe height $h$ (amplitude-phase-distance curves) and are then converted into values of the force quadratures\cite{Paulo2002}. In contrast to FM-AFM, the oscillation amplitude is free to change during the measurement and thus the AM-AFM measurement path in the $h$-$A$-plane is more complicated. The path shown in figure \ref{fig:F-model-maps} was obtained by simulating the AM-AFM tip dynamics with cvode. In the simulations we used the same cantilever and force parameters as in the previous section with a drive signal at only one frequency of $\omega_{d}=300\ \mathrm{kHz}$. As is often the case with AM-AFM, the amplitude-phase-distance curve exhibits an abrupt amplitude jump due to the existence of multiple oscillation states\cite{Garcia1999}. This instability is frequently observed in experiments, and it makes the reconstruction of tip-surface forces rather difficult.

In contrast to FM-AFM and AM-AFM, ImAFM allows for a measurement of $F_I$ and $F_Q$ at fixed static probe height, along a straight path parallel to the $A$-axis in the $h$-$A$ plane as shown in figure \ref{fig:F-model-maps} for the simulation of the previous section.  Each of these three measurement techniques probes the tip-surface interaction along a different path in the $h$-$A$ plane.  With ImAFM however, the measurement can be rapidly preformed in each point of an image, while scanning with normal speed \cite{Platz2008} allowing for unprecedented ability to analyse the tip-surface force while imaging.  The ImAFM spectral data, which is concentrated to a narrow band near resonance, is a complete representation of the measurable tip motion, because there is only noise outside this narrow frequency band.   Thus the method optimally extracts the signal for compact storage and further analysis.  

We note that the ImAFM path provides an equivalent amount of information as frequency-shift-distance or amplitude-phase-distance curves. This implies that for a single scan ImAFM image information equivalent to a frequency-shift-distance curve or an amplitude-phase-distance curve is available in every image point. Moreover, the ImAFM measurement does not suffer from amplitude jumps since the stiffness of the cantilever resonance prevents big amplitude changes from one single oscillation cycle to the next single oscillation in the beat tip motion.

\subsection{ImAFM approach measurements}

It is possible to acquire maps of $F_I$ and $F_Q$ in the full $h$-$A$ plane with a protocol we call ImAFM approach measurements. Similar to the measurement of  frequency-shift or amplitude-phase-curves, the static probe height above the surface is varied by slowly extending the z-piezo toward the surface.   However, in contrast to FM-AFM and AM-AFM measurements, the oscillation amplitude is rapidly modulated as the probe slowly approaches the surface. Because the height variation is much slower  (order of seconds) than the amplitude modulation (order of milliseconds), the probe height can be considered to be constant during each amplitude modulation. In this case each amplitude modulation reveals the amplitude dependence of $F_I$ and $F_Q$ at a constant probe height.   From the different probe heights $F_I$ and $F_Q$ can be reconstructed can be reconstructed in the full $h$-$A$ plane. With FM-AFM or AM-AFM such a measurement would require much longer measurement time since multiple surface approaches with different amplitudes would be required. With ImAFM all the data is acquired during a single surface approach.

\begin{figure}
\centering{}\includegraphics{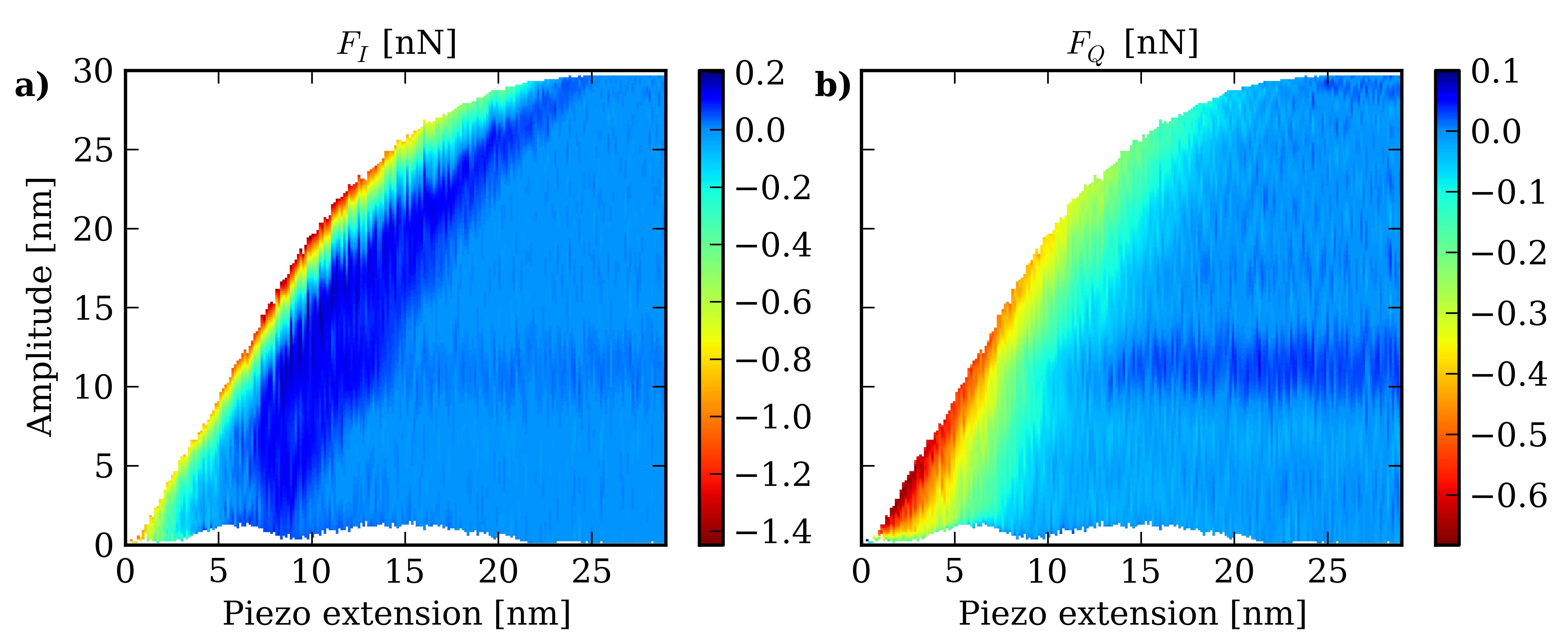}\caption{$F_I(h,A)$ and $F_Q(h,A)$ maps reconstructed from an ImAFM approach measurements on a PS surface. The z-piezo extension corresponds to a relative change of the probe height $h$ above the sample surface. \label{fig:f-maps-exp}}
\end{figure}

We use ImAFM approach curves to reconstruct $F_I$ and $F_Q$ maps on a Polystyrene (PS) polymer surface. We perform a slow surface approach and from the acquired data we reconstruct the $F_I$ and $F_Q$ maps shown in figure \ref{fig:f-maps-exp}. On the $h$-axis we show the piezo extension since the absolute probe height cannot be defined unambiguously in an experiment. The areas in the $h$-$A$ plane that were not explored are displayed as white areas. 

The boundary to the white area in the upper part of the plots represents the maximum oscillation achievable for the fixed drive power. The boundary to the white area in the lower part of the plot corresponds to the minimum oscillation amplitude during one modulation period (one beat). This lower boundary shows interesting variations with the piezo extension. Further away from the surface at larger piezo extension, only positive values of $F_I$ are achieved, which corresponds to the tip oscillating in a region where the net conservative force is purely attractive. As the surface is further approached, $F_I$ also takes on negative values, when the net force becomes repulsive. At probe heights between 7 and 13 nm the strongest repulsive force is experienced. Around $h=$13 nm and below, the attractive region vanishes, which may be the result of a change in the cantilever dynamics due to relatively long interaction time, or a change in the hydrodynamic damping forces due to the surrounding air close the sample surface. One should also note that at this piezo extension the minimum oscillation amplitude begins to increase again. A possible artifact of the measurement method may result  in this low amplitude region, if the motion spectrum is no longer confined to a narrow frequency band, as assumed in the analysis.

The map of $F_Q$ characterizes the dissipative interaction between tip and sample. The dissipative tip-surface force can be much more complex than the conservative part of the interaction since dissipative forces do not only depend on the instantaneous tip position as with the conservative force. The $F_Q$ map can provide detailed insight into the nature of the dissipative interaction since the full dependence of $F_Q$ on probe height $h$ and oscillation amplitude $A$ is measured. The $F_Q$ map shows a lower level of force than the $F_I$ map and it therefore appears more noisy, because the tip-surface interaction is predominately conservative. The small positive values of $F_Q$ which occur far from the surface would imply that the tip gained energy from the interaction with the surface. This may be an artifact, but another possible explanation is some sort of hydrodynamic mode above the surface\cite{Fontaine1997}. In both the attractive and repulsive region of the in-phase force $F_I$ , the quadrature force $F_Q$ is predominantly negative and it decreases as the surface is indented, corresponding to a increasingly dissipative tip-surface interaction. However, the maximum dissipation does not coincide with the maximum repulsive conservative force, and the energy dissipation is largest at peak amplitude for piezo extensions between 2 and 6 nm. Another interesting feature of the $F_I$ map is the fine structure in the contact region, between $h=$ 10 and 20 nm piezo extension.  These  small step-like changes of conservative force are not present in the smooth force model function, and could be indication that the dissipative forces are resulting in small, irreversible modifications of the sample surface.

\section{Conclusion}

We presented a physical interpretation of tip motion when described by a narrow-band frequency comb in ImAFM. We showed by separation of time scales that the time domain signal of a narrow-band frequency comb is completely characterized by a complex-valued envelope function and a rapidly oscillating term. The application of this time domain picture to ImAFM allows for the reconstruction of two force quadratures $F_I$ and $F_Q$ as functions of the oscillation amplitude $A$. The quantities $F_I$ and $F_Q$ can be considered as two-dimensional functions, depending on the both probe height and the oscillation amplitude. Within this framework we find a connection between frequency-shift-distance curves in FM-AFM, amplitude-phase-distance-curves in AM-AFM, and ImAFM measurements. Moreover, we introduced ImAFM approach measurements which allow for a rapid and complete reconstruction of $F_I$ and $F_Q$ in the full $h$-$A$ plane, providing detailed insight into the interaction between tip and surface. We demonstrated the reconstruction of $F_I$ and $F_Q$ maps experimentally on a PS polymer surface. We hope that the physical interpretation of narrow-band dynamic AFM presented here, will inspire new force spectroscopy methods in the future which take advantage of the high signal-to-noise ratio and the high acquisition speed of ImAFM.

\section{Experimental}

The PS sample was spin-cast from Toluene solution on a silicon oxide substrate. Both PS ($M_{w}=280\ \mathrm{kDa})$ and Toluene were obtained from Sigma-Aldrich and used as purchased. The measurements where performed with a Veeco Multimode II and a Budget Sensor BS300Al-G cantilever with a resonance frequency of $f_{0}=311.838\ \mathrm{kHz}$, a quality factor of $Q=539.9$ and a stiffness of $k_c=29.5\ \mathrm{N/m}$ which was determined by thermal calibration\cite{Higgins2006}. We choose the two drive frequencies $f_{1}=311.585\ \mathrm{kHz}$ and $f_{2}=312.085\ \mathrm{kHz}$ symmetrically around the resonance frequency and the drive strengths such that the free oscillation amplitude is modulated between 0.0 and 29.7 nm. The probe height is changed with a speed of 5.0 nm/s.

\end{document}